\documentclass{pasj00}
\twocolumn
\usepackage{times}

\begin{document}
\SetRunningHead{K. Yabe et al.}{FastSound: the mass-metallicity relation of galaxies}

\title{The Subaru FMOS Galaxy Redshift Survey (FastSound). III. The mass-metallicity relation and the fundamental metallicity relation at $z\sim1.4$\thanks{Based on data collected at Subaru Telescope, which is operated by the National Astronomical Observatory of Japan.}}

\author{Kiyoto \textsc{Yabe}\altaffilmark{1, 2}, Kouji \textsc{Ohta}\altaffilmark{3}, Masayuki \textsc{Akiyama}\altaffilmark{4}, Andrew \textsc{Bunker}\altaffilmark{2,5}, Gavin \textsc{Dalton}\altaffilmark{5,6}, Richard \textsc{Ellis}\altaffilmark{7}, Karl \textsc{Glazebrook}\altaffilmark{8}, Tomotsugu \textsc{Goto}\altaffilmark{9}, Masatoshi \textsc{Imanishi}\altaffilmark{10}, Fumihide \textsc{Iwamuro}\altaffilmark{3}, Hiroyuki \textsc{Okada}\altaffilmark{11}, Ikkoh \textsc{Shimizu}\altaffilmark{11}, Naruhisa \textsc{Takato}\altaffilmark{10}, Naoyuki \textsc{Tamura}\altaffilmark{2}, Motonari \textsc{Tonegawa}\altaffilmark{11}, \& Tomonori \textsc{Totani}\altaffilmark{11}}

\altaffiltext{1}{Division of Optical and IR Astronomy, National Astronomical Observatory of Japan, 2-21-1, Osawa, Mitaka, 181-8588, Japan}
\email{kiyoto.yabe@nao.ac.jp, kiyoto.yabe@ipmu.jp}
\altaffiltext{2}{Kavli Institute for the Physics and Mathematics of the Universe (WPI), The University of Tokyo, 5-1-5 Kashiwanoha, Kashiwa, Chiba 277-8583, Japan}
\altaffiltext{3}{Department of Astronomy, Kyoto University, Kitashirakawa Oiwake-cho, Sakyo-ku, Kyoto, 808-8588, Japan}
\altaffiltext{4}{Astronomical Institute, Tohoku University, 6-3 Aramaki, Aoba-ku, Sendai, 980-8578, Japan}
\altaffiltext{5}{Department of Physics, University of Oxford, Denys Wilkinson Building, Keble Road, Oxford OX1 3RH, UK}
\altaffiltext{6}{STFC Rutherford Appleton Laboratory, Chilton, Didcot, Oxfordshire OX11 0QX, UK}
\altaffiltext{7}{Department of Astrophysics, California Institute of Technology, MS 249-17, Pasadena, CA 91125, USA}
\altaffiltext{8}{Centre for Astrophysics and Supercomputing, Swinburne University of Technology, P.O. Box 218, Hawthorn, VIC 3122, Australia}
\altaffiltext{9}{Department of Physics, National Tsing Hua University, 101 Section 2 Kuang Fu Road, Hsinchu 30013, Taiwan}
\altaffiltext{10}{Subaru Telescope, 650 North A'ohoku Place, Hilo, Hawaii, 96720, USA}
\altaffiltext{11}{Department of Astronomy, The University of Tokyo, 7-3-1 Hongo, Bunkyo-ku, Tokyo 113-0033, Japan}


%

\KeyWords{galaxies: evolution --- galaxies: abundances --- galaxies: high-redshift --- galaxies: ISM} 

\maketitle

\begin{abstract}
We present the results from a large near-infrared spectroscopic survey with Subaru/FMOS (\textit{FastSound}) consisting of $\sim$ 4,000 galaxies at $z\sim1.4$ with significant H$\alpha$ detection. We measure the gas-phase metallicity from the [N~{\sc ii}]$\lambda$6583/H$\alpha$ emission line ratio of the composite spectra in various stellar mass and star-formation rate bins. The resulting mass-metallicity relation generally agrees with previous studies obtained in a similar redshift range to that of our sample. No clear dependence of the mass-metallicity relation with star-formation rate is found. Our result at $z\sim1.4$ is roughly in agreement with the fundamental metallicity relation at $z\sim0.1$ with fiber aperture corrected star-formation rate. We detect significant [S~{\sc ii}]$\lambda\lambda$6716,6731 emission lines from the composite spectra. The electron density estimated from the [S~{\sc ii}]$\lambda\lambda$6716,6731 line ratio ranges from 10 -- 500 cm$^{-3}$, which generally agrees with that of local galaxies. On the other hand, the distribution of our sample on [N~{\sc ii}]$\lambda$6583/H$\alpha$ vs. [S~{\sc ii}]$\lambda\lambda$6716,6731/H$\alpha$ is different from that found locally. We estimate the nitrogen-to-oxygen abundance ratio (N/O) from the N2S2 index, and find that the N/O in galaxies at $z\sim1.4$ is significantly higher than the local values at a fixed metallicity and stellar mass. The metallicity at $z\sim1.4$ recalculated with this N/O enhancement taken into account decreases by 0.1 -- 0.2 dex. The resulting metallicity is lower than the local fundamental metallicity relation.
\end{abstract}

\section{Introduction\label{sec:introduction}}
The metallicity in the gas-phase (hereafter, metallicity) is one of the key parameters required for understanding galaxy formation and evolution. Heavy elements produced in stars through the star-formation activity are returned into the inter-stellar medium (ISM), which is also regulated by gas infall and outflows. Thus, the metallicity can be a tracer of the past star-formation history including the gas flow history (e.g.,\cite{Tinsley:1980}). Enormous efforts on studying the correlation between the stellar mass and metallicity (hereafter mass-metallicity relation) over a wide dynamic range has been made at $z\lesssim0.3$ (e.g., \cite{Lequeux:1979}, \cite{Tremonti:2004p4119}, \cite{Gallazzi:2005}, \cite{Lee:2006}, \cite{Kewley:2008}, and \cite{Andrews:2013}), at intermediate redshift of $0.3 \lesssim z \lesssim 1$ (e.g., \cite{Savaglio:2005}, \cite{Rodrigues:2008}, \cite{Lamareille:2009}, \cite{Zahid:2013}, \cite{Ly:2014}), and at higher redshifts, $z\gtrsim1$ (e.g., \cite{Erb:2006p4143}, \cite{Maiolino:2008}, \cite{Mannucci:2009}, \cite{Hayashi:2009}, \cite{Yabe:2012}, \cite{Belli:2013}, \cite{Henry:2013}, \cite{Yabe:2014}, \cite{Zahid:2014b}, \cite{Cullen:2014}, \cite{Steidel:2014}, and \cite{Sanders:2015}). The compilation of these studies shows that the mass-metallicity relation evolves with redshift such that the metallicity increases gradually with decreasing redshift at a fixed stellar mass (\cite{Maiolino:2008}, and \cite{Zahid:2013}).

The intrinsic scatter of the mass-metallicity relation and its dependence of the star-formation activity has also been investigated (e.g., \cite{Tremonti:2004p4119}, \cite{Ellison:2008}, \cite{Lara-Lopez:2010b}, \cite{Mannucci:2010p8026}, \cite{Yates:2012}, and \cite{Andrews:2013}). A fundamental relation between stellar mass, metallicity, and star-formation rate (SFR) for Sloan Digital Sky Survey (SDSS) galaxies at $z\sim0.1$ has been proposed by \citet{Mannucci:2010p8026}, with an extension to galaxies of lower stellar mass (\cite{Mannucci:2011}, \cite{Hunt:2012}, and \cite{Zahid:2012}). According to this fundamental metallicity relation (hereafter, FMR), galaxies with higher SFR show lower metallicity at a fixed stellar mass. The evolution of the mass-metallicity relation can be explained by the evolution of the SFR of galaxies \citep{Mannucci:2010p8026}. The presence of a FMR and its origin has, however, been argued in many studies (e.g., \cite{Mannucci:2010p8026}, \cite{Bothwell:2013}, \cite{Forbes:2014}, and \cite{Zahid:2014a}).

Recently, the presence of the FMR at higher redshift has been explored in various studies (e.g., \cite{Cresci:2012}, \cite{Yabe:2012}, \cite{Vergani:2012}, \cite{Wuyts:2012}, \cite{Stott:2013}, \cite{Yabe:2014}, \cite{Zahid:2014b}, \cite{Stott:2014}, \cite{Divoy:2014}, and \cite{Steidel:2014}). The dependence of the mass-metallicity relation on the SFR at high redshift, however, still remains unclear. Some studies suggest a dependence of the mass-metallicity relation on SFR or specific SFR within their samples (\cite{Yabe:2012}, \cite{Stott:2013}, \cite{Stott:2014}, and \cite{Zahid:2014b}), but others do not show such a clear dependence (\cite{Yabe:2014}, \cite{Steidel:2014}, and \cite{Sanders:2015}). This disagreement may be partly due to the smallness of the samples, as the number of galaxies in each study is still limited up to $\sim$300. Furthermore, no clear SFR dependence is due to the narrow range of the SFR (see discussions by \cite{Stott:2013, Yabe:2014}). Furthermore, some studies suggest offsets at high redshift from the local FMR. By using their sample at $z=$ 0.84 -- 1.6, \citet{Yabe:2012}, \citet{Stott:2013}, and \citet{Yabe:2014} reported that the metallicity is generally higher than the extrapolation of the local FMR to the lower values of $\mu$ ($\equiv \textrm{log}(M_{*})-\alpha\textrm{log}(SFR)$), where $\alpha$ is the projection parameter of the FMR (see \cite{Mannucci:2010p8026}).  On the other hand, \citet{Cullen:2014} also reported the deviation from the local FMR at $z\sim2-4$ with R23 metallicity indicator, but they suggest a counter trend; the metallicity is lower than the local relation. A Larger sample spanning a wider range of parameter space is needed to reveal the FMR and other parameter dependence of the mass-metallicity relation at high redshift.

The \textit{FastSound} project \citep{Tonegawa:2015b} is a cosmological redshift survey to measure redshifts of several thousand star-forming galaxies at $z$ = 1.18 -- 1.54 in the total area of $\sim$20 deg$^{2}$ in the Canada-France-Hawaii Telescope Legacy Survey (CFHTLS) wide fields by using the H$\alpha$ emission lines obtained with Fiber Multi-Object Spectrograph (FMOS) on the Subaru Telescope. Although the main science goal of the survey is to measure the redshift space distortion in this redshift range, the survey also provides for ancillary science cases with a large near-infrared (NIR) spectroscopic sample of H$\alpha$ emitting galaxies at $z>1$. The targets in this survey are galaxies which are relatively bright in H$\alpha$ emission line, observed in a short exposure time, i.e., the target galaxies have moderately high SFR. In this work, by using an unprecedentedly large NIR spectroscopic sample, we examine the mass-metallicity relation of star-forming galaxies and explore the high-SFR end of the FMR.

Throughout this paper, we use the following cosmological parameters: $H_{0}=70$ km s$^{-1}$ Mpc$^{-1}$, $\Omega_{\Lambda}=0.7$, and $\Omega_{m}=0.3$. All magnitudes given in this paper are in the AB magnitude system. We use the Salpeter initial mass function (IMF).

\section{Sample Selection and Observations\label{sec:sample}}
\begin{figure}
\begin{center}
\includegraphics[angle=270,scale=0.50]{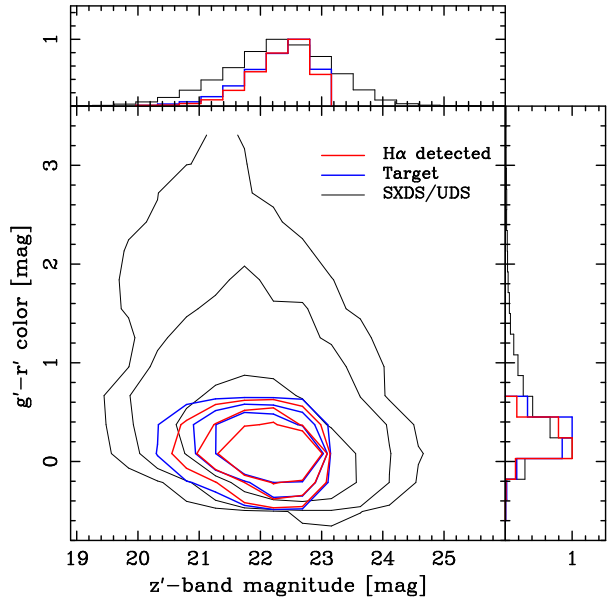}
\caption{The distribution of $z'$-band magnitude and $g'-r'$ color for target sample (blue contour), H$\alpha$ detected sample (red contour), and the  K-selected photometric redshift sample at $1.2<z_{ph}<1.5$ in the SXDS/UDS field (black contour). The contour levels are 68\%, 90\%, and 97\%, respectively. In the top and right panels, the peak-normalized histograms of the magnitude and color are shown for each sample.\label{fig:MagCol}}
\end{center}
\end{figure}

The target sample in the \textit{FastSound} survey is selected by photometric redshift and the predicted H$\alpha$ flux based on the SFR and color excess from SED fitting with $u^{*}$, $g'$, $r'$, $i'$, and $z'$ photometric data of the four CFHTLS wide fields (W1, W2, W3, and W4). The photometric redshifts of the galaxies are derived using \textit{LePhare} \citep{Ilbert:2006}. As a target sample, galaxies with $1.1 < z_{ph} < 1.6$, $20.0<z'<23.0$ mag, and $g'-r'<0.55$ mag are selected. Furthermore, an additional criterion of an expected H$\alpha$ flux of $\gtrsim 1.0 \times 10^{-16}$ erg s$^{-1}$ cm$^{-2}$ is applied. The magnitude, color, and the expected H$\alpha$ flux cut for the target selection are made aiming to enhance the detection rate in the survey. It should be noted that some selection biases are raised by the expected H$\alpha$ flux cut, though the effect is marginal according to the previous studies (\cite{Yabe:2012}, and \cite{Yabe:2014}). The detailed descriptions on the sample selection are presented by \citet{Tonegawa:2015b}. In figure \ref{fig:MagCol}, the distribution of our target sample in $z'$-band magnitude vs. $g'-r'$ color diagram is shown with more completed sample at $1.2 < z_{ph} < 1.5$ in the SXDS/UDS field by \citet{Yabe:2014}. The peak-normalized histograms show that our target sample is relatively biased toward brighter magnitude and bluer color compared to the SXDS/UDS sample, because of our magnitude and color selection. Dusty and/or old galaxies may be left out of our sample selection.

The observations were carried out in 40 nights from September 2011 to January 2014 by using Subaru/FMOS, which is a fiber-fed multi-object NIR spectrograph with 400 fibers (1.2\arcsec\ diameter) at the prime focus of the Subaru Telescope (30\arcmin\ diameter). Details of the instrument are presented by   \citet{Kimura:2010}. We use the spectrograph setting of the high resolution mode (H-short prime; $R\sim2400$), covering the wavelength range from 1.45 $\mu$m to 1.67 $\mu$m. We used the normal beam switching mode during the observations, in which all the fibers are available for the target objects (up to $\sim400$). In the typical observational process of this survey, after the initial fiber configuration, a set of on-target and off-target frame with exposure time of 15 minutes each is taken. The typical beam-switch offset is 10\arcsec\ -- 15\arcsec\ from the target. After re-configuration of the fiber positions, another set of the on- and off-frame is taken in the same manner. The total on-source exposure time is 30 minutes for each object. The seeing size, which was monitored with the optical sky camera during the observations, ranged from $\sim0.5\arcsec$ to $\sim1.7\arcsec$ with a median value of $1.08\arcsec$. We observed 118  field-of-views in the four CFHTLS wide fields ($\sim$20 deg$^{2}$), covering 38,615 targets in total.

The basic data reductions are carried out by using the standard FMOS pipeline \textit{FIBRE-pac} \citep{Iwamuro:2011p18754} \footnote{Since the wavelength calibration of the FMOS data is carried out with vacuum wavelength, we use the vacuum wavelength of each emission line in the analysis of this work. Throughout this paper, however, we note the wavelength of each line in air, such as [N~{\sc ii}]$\lambda$6583, [S~{\sc ii}]$\lambda\lambda$6716,6731, etc.}. The emission line detection from the extracted 1D spectra is carried out by using an automated emission line detection software (FIELD), which was developed and tested for the \textit{FastSound} survey \citep{Tonegawa:2015}. We detect H$\alpha$ emission lines from 9,803 galaxies with signal-to-noise ratio (S/N)$\geq$3.0,   which include 4,139 and 2,740 galaxies with S/N$\geq$4.0 and 5.0, respectively. According to the analysis by \citet{Tonegawa:2014} using inverted images, the contamination rate due to false detections is $\sim 5 \%$ for a S/N cut of 4.0. In this work, we use the S/N threshold of 4.0 for H$\alpha$ detection. The spectroscopic redshifts of the detected objects range from 1.18 to 1.54 with the average (median) value of 1.362 (1.356). The redshift distribution is presented by \citet{Okada:2015}. The observed line width (FWHM$^{\textrm{\footnotesize{obs}}}$) ranges from $\sim100$ to $\sim1000$ km s$^{-1}$ with the average (median) value of 232 (212) km s$^{-1}$. After deconvolving the spectral resolution ($R\sim2400$) by subtracting in quadrature the FWHM measured from unresolved sky lines, the average (median) intrinsic line width (FWHM$^{\textrm{\footnotesize{int}}}$) is 203 (187) km s$^{-1}$. Figure \ref{fig:MagCol} shows that  the H$\alpha$ detected sample has similar $z'$-band magnitude and $g'-r'$ color range to those of the target sample.

Since our observations targeting the H$\alpha$ detection do not cover the $J$-band, we have no information on H$\beta$ and [O~{\sc iii}]$\lambda\lambda$4959,5007. The emission line ratio diagnostics using [N~{\sc ii}]$\lambda$6583/H$\alpha$ and [O~{\sc iii}]$\lambda$5007/H$\beta$ \citep{Baldwin:1981p15095} and the recent mass-excitation diagnostics using stellar mass and [O~{\sc iii}]$\lambda$5007/H$\beta$ line ratio \citep{Juneau:2014} cannot be used. According to diagnostics by \citet{Kewley:2006}, in the range of the obtained [N~{\sc ii}]$\lambda$6583/H$\alpha$ in this work, a spectrum is identified as a star-forming region if log([O~{\sc iii}]$\lambda$5007/H$\beta$) is $\lesssim0.7-0.8$. In the stacked spectrum of the whole sample, which will be described in detail in the next section, [O~{\sc i}]$\lambda$6300 emission line is detected with S/N$\sim$20. According to \citet{Kewley:2006}, a measured log([O~{\sc i}]$\lambda$6300/H$\alpha$) of -1.55 is identified with star-forming galaxies if log([O~{\sc iii}]$\lambda$5007/H$\beta$) is $\lesssim0.6$. In the study by \citet{Yabe:2014}, most of the sample galaxies in the redshift range of our sample show log([O~{\sc iii}]$\lambda$5007/H$\beta$)$\lesssim0.6$. In any event, active galactic nuclei (AGNs) with these low [N~{\sc ii}]$\lambda$6583/H$\alpha$ and [O~{\sc i}]$\lambda$6300/H$\alpha$ are very rare in the local universe. Furthermore, as we will discuss in subsection \ref{subsec:noo}, the [N~{\sc ii}]$\lambda$6583/H$\alpha$ and [S~{\sc ii}]$\lambda\lambda$6716,6731/H$\alpha$ line ratios indicate that our sample does not suffer severely from AGN contamination. As we mentioned above, most of the line widths of our sample are $\lesssim$1000 km s$^{-1}$, which means that our sample is not severely contaminated by broad line AGNs. In this work, we exclude only a marginal fraction ($\sim$1.1 \%) of broad line emission galaxies (FWHM$^{\textrm{\footnotesize{obs}}}$ $>1000$ km s$^{-1}$) from the analysis below. Thus, the total number of galaxies used in this work after excluding the possible broad AGN candidates is 4,094.

\begin{figure}
\begin{center}
\includegraphics[angle=270,scale=0.50]{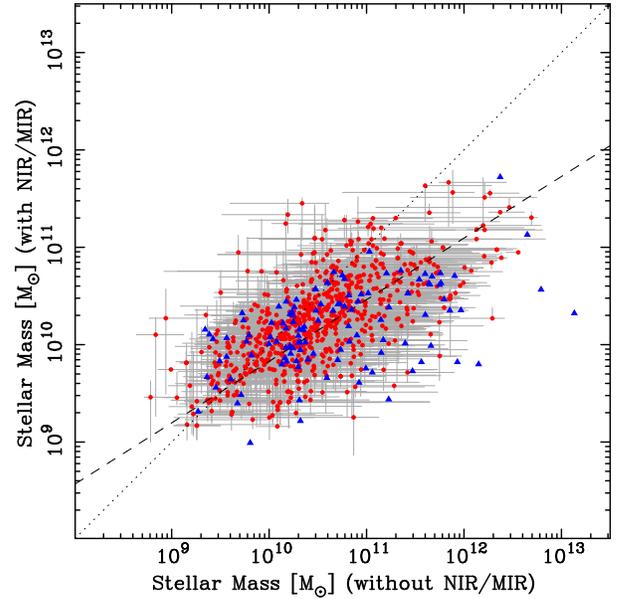}
\caption{Comparison between stellar mass derived with NIR/MIR photometry and that without NIR/MIR photometry. Red filled circles are objects with UKIDSS/DXS data and blue filled triangles are objects with UKIDSS/UDS data. Dotted line shows an equivalent line and dashed line indicates the linear regression line. \label{fig:MsComp}}
\end{center}
\end{figure}

\begin{figure}
\begin{center}
\includegraphics[angle=270,scale=0.50]{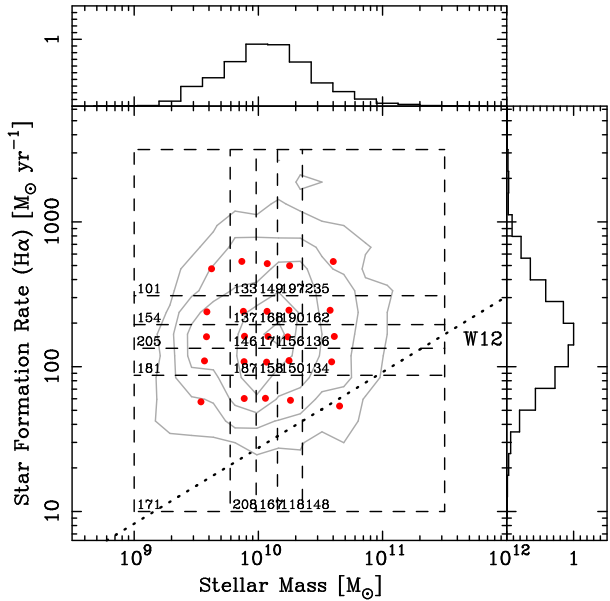}
\caption{The distribution of the stellar mass and the SFR of our sample. Contours for 96\%, 85\%, 66\%, 24\%,  and 4\% level are shown in gray curves. The average value in each stellar mass and SFR bin (indicated by horizontal and vertical dashed lines) is shown by a red circle. The number of galaxies in each bin is also presented at the bottom left corner in each section. The dotted line is the galaxy main sequence at $z\sim1.4$ by \citet{Whitaker:2012}. In the top and right panels, the peak-normalized distribution of stellar mass and SFR is shown, respectively. \label{fig:MsSFR}}
\end{center}
\end{figure}

\section{Measurements of stellar mass, star-formation rate, and metallicity\label{sec:methods}}
\subsection{Stellar masses \label{sec:methods:mass}}
The stellar masses used in the present analysis are derived from multi-wavelength spectral energy distribution (SED) fitting using photometric data from the rest-frame ultraviolet (UV) to mid-infrared (MIR). The initial catalogue of our sample with optical photometry is first cross-matched to the UKIRT (United Kingdom Infrared Telescope) Infrared Deep Sky Survey (UKIDSS) Deep Extragalactic Survey (DXS) catalogue derived from UKIRT Wide Field Camera (WFCAM) imaging data in the $J$- and $K$-band near-infrared (NIR) bands \citep{Lawrence:2007}, which covers part of CFHTLS W1 and W4. In total, 487 objects out of 701 objects are detected in the $J$ and/or $K$ bands. The total magnitude of the detected objects in the NIR with S/N$\geq$5 ranges from $J\sim$ 20.0 -- 23.0 mag and $K\sim$ 19.5 -- 22.8 mag. Our sample is also cross-matched to the Spitzer Enhanced Imaging Products\footnote{http://irsa.ipac.caltech.edu/data/SPITZER/Enhanced/SEIP/overview.html} (SEIP) catalogue derived from Spitzer Infrared Array Camera (IRAC) 3.6$\mu$m (ch.1) and $4.5\mu$m (ch.2) MIR data, where combined images are constructed from several earlier programs. In total, 422 objects of our sample are detected in ch.1 and/or ch.2 band. The total magnitude of the detected objects in MIR with S/N$\geq$5 ranges from ch.1 $\sim$ 19.0 -- 22.6 mag and ch.2 $\sim$ 18.8 -- 22.8 mag. 791 objects ($\sim20\%$ of the total sample) are detected in at least one of the NIR to MIR bands.

SED fitting was undertaken using the \textit{SEDfit} code \citep{Sawicki:2012} as described by \citet{Yabe:2014}. Here, we assume the Salpeter IMF (0.1 -- 100 M$_{\odot}$) and the Calzetti extinction law. For objects without NIR or MIR data, stellar masses are estimated using only optical data from the $u^{*}$- to $z'$-bands. To evaluate the uncertainties arising from the lack of NIR and/or MIR (hereafter, simply NIR/MIR) data, we compare stellar masses derived with and without NIR/MIR data by using the subsample with both optical and NIR/MIR photometry in figure \ref{fig:MsComp}. Although the stellar masses derived without using the NIR/MIR data are larger by up to $\sim1$ dex in the most massive end, the agreement is reasonably good in the mass range from $\sim10^{9}$ to  a few $\times$ $10^{10}$ M$_{\solar}$, in which most of our sample galaxies are. We discuss the possible effect of the uncertainty in the stellar mass derivation on results in section \ref{subsec:MZR}. Using a linear function to correct this bias, we find: $\textrm{log}(M_{*}^{\textrm{\scriptsize{with NIR/MIR}}})=0.63\times \textrm{log}(M_{*}^{\textrm{\scriptsize{without NIR/MIR}}})+3.50$, dashed line in figure \ref{fig:MsComp}. with a scatter corresponding to $\sigma=0.22$ dex. In the following analysis, if the stellar mass can only be derived from optical data, we correct it applying this relation.

The comparison of the stellar mass derived with and without NIR/MIR data in figure \ref{fig:MsComp} could be biased if the nature of the SED of the NIR/MIR undetected sample differs from that where NIR/MIR detection is available. This bias could be critical for the less massive portion of figure \ref{fig:MsComp}. To address this, we check the stellar masses for a subset of our sample in the UKIDSS Ultra Deep Survey (UDS DR8; \cite{Lawrence:2007}) field, where deeper NIR data is available. The UDS field covers part of CFHTLS W1 but is not overlapped with DXS field. In this deeper NIR sample, 115 out of 140 of our objects are detected down to $\sim$23.6 mag with S/N$\gtrsim$10. The stellar masses derived with the deeper NIR data are compared to that without the NIR data. Figure \ref{fig:MsComp} shows that the comparison using deeper UDS data remains consistent with the regression line derived using the DXS data.

\subsection{Star-formation rates \label{sec:methods:sfr}}
The star formation rate (SFR) is estimated from the extinction corrected H$\alpha$ luminosity according to the methodology discussed by \citet{Yabe:2012} and \citet{Yabe:2014}. The color excess for the stellar continuum is derived from $u^{*} - r'$ color, corresponding to the rest-frame UV for the redshift range of our sample, with an extinction law by \citet{Calzetti:2000}. Based on the prescription by \citet{CidFernandes:2005}, we correct the broad-band color excess to that for nebular emission, applying a factor of $\sim2$. We use the relation by \citet{Kennicutt:1998} to infer the SFR from the H$\alpha$ luminosity. Following the study by \citet{Tonegawa:2014} using a sample similar to the present one, we applied a fiber aperture correction of 0.5. The uncertainties of our flux estimates will be described in \citet{Okada:2015}. The SFRs derived from our H$\alpha$ luminosities agree with those inferred from the dust-corrected rest-frame UV luminosity density with a scatter of only $\sigma=0.26$ dex. The distribution of stellar masses and SFRs of our sample is presented in figure \ref{fig:MsSFR} and summarized in table \ref{tab1}. It is worth noting that our sample mostly comprises galaxies with higher SFRs  compared to the `main sequence' at $z\sim1.4$ \citep{Whitaker:2012}.

\begin{table}
\caption{Stellar mass and SFR of each bin in our sample. \label{tab1}}
\begin{center}
\begin{tabular}{cccc}
\hline
Bin & Num.$^{a}$ & log($M_{*}$)$^{b}$ & log(SFR)$^{b}$ \\ & & [M$_{\odot}$] & [M$_{\odot}$ yr$^{-1}$] \\
\hline
1 & 171 & 9.55 & 1.81\\
2 & 208 & 9.90 & 1.81\\
3 & 167 & 10.05 & 1.82\\
4 & 118 & 10.26 & 1.83\\
5 & 148 & 10.61 & 1.81\\
6 & 181 & 9.57 & 2.05\\
7 & 187 & 9.89 & 2.03\\
8 & 158 & 10.06 & 2.03\\
9 & 150 & 10.24 & 2.04\\
10 & 134 & 10.53 & 2.03\\
11 & 205 & 9.61 & 2.21\\
12 & 146 & 9.90 & 2.21\\
13 & 171 & 10.08 & 2.21\\
14 & 156 & 10.23 & 2.21\\
15 & 136 & 10.55 & 2.21\\
16 & 154 & 9.61 & 2.38\\
17 & 137 & 9.89 & 2.38\\
18 & 168 & 10.07 & 2.37\\
19 & 190 & 10.25 & 2.39\\
20 & 162 & 10.52 & 2.39\\
21 & 101 & 9.65 & 2.62\\
22 & 133 & 9.86 & 2.71\\
23 & 149 & 10.08 & 2.65\\
24 & 197 & 10.25 & 2.65\\
25 & 235 & 10.55 & 2.67\\
\hline
\end{tabular}
\\Note -- (a) the number of galaxies in each stellar mass and SFR bin. (b) median values in each bin. 
\end{center}
\end{table}

\subsection{Metallicities \label{sec:methods:metallicity}}
The metallicities in our analysis are derived from the [N~{\sc ii}]/H$\alpha$ line ratio according the relation introduced by \citet{Pettini:2004p7356}:
\begin{equation}
\textrm{12+log(O/H)}=8.90+0.57 \times \textrm{N2},
\end{equation}
where N2 $=$ log([N~{\sc ii}]$\lambda$6583/H$\alpha$). Although it is known that the N2 index saturates above solar metallicity, most of our sample has sub-solar metallicity and the effect of this saturation will be small. In our analysis, the errors do not include any possible intrinsic error in the calibration ($\sim$0.2 dex; \cite{Pettini:2004p7356}) or potential systematic differences between the various metallicity indicators (up to $\sim$0.7 dex; \cite{Kewley:2008}).

\begin{figure*}[]
\begin{center}
\includegraphics[angle=270, scale=1.00]{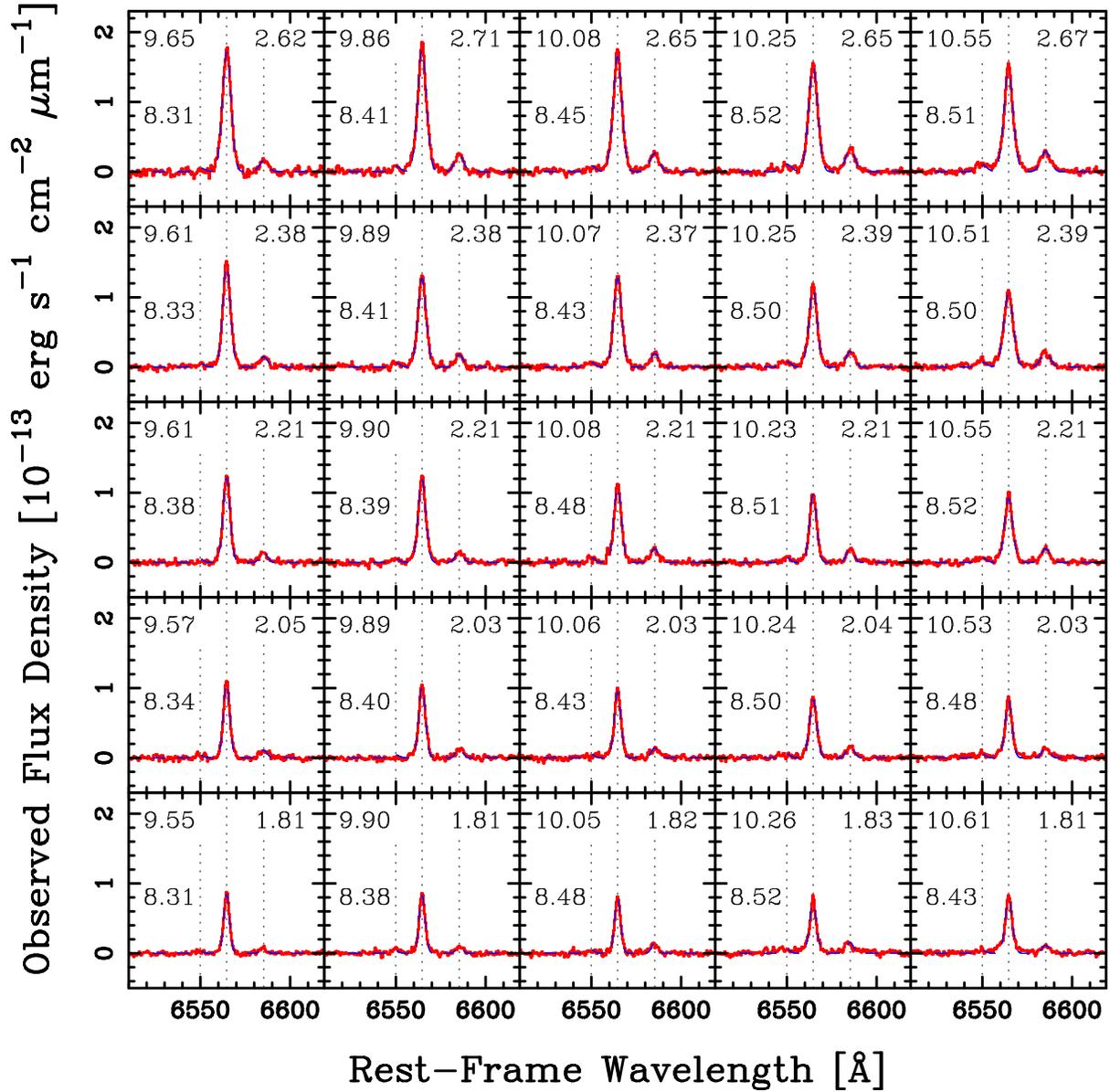}
\caption{A composite spectrum of our sample at $z\sim1.4$ in each stellar mass and SFR bin. The stellar mass increases from left to right and the SFR increases from bottom to top. The observed spectra and the best-fit model spectra are shown by red solid lines and blue dashed lines, respectively. The positions of H$\alpha$ and [N~{\sc ii}]$\lambda\lambda$6548,6583 are indicated by vertical dotted lines. The median values of log($M_{*}$[M$_{\solar}$]), log(SFR[M$_{\solar}$ yr$^{-1}$]), and 12+log(O/H) are shown at the top-left, top-right, and middle-left part of each panel, respectively. \label{fig:SpectraHaN2}}
\end{center}
\end{figure*}

[N~{\sc ii}]$\lambda$6583 emission lines from 224 objects are detected with S/N $\ge$ 3, while a large part of our sample shows no detection. In order to measure the average metallicity and establish the mass-metallicity relation, we apply a spectral stacking analysis follow the method described by \citet{Yabe:2012} and \citet{Yabe:2014}. We construct a weighted-mean composite spectrum in each stellar mass and SFR bin. Since the spectra in this study were all obtained in FMOS' high-resolution mode, no correction for the obscuring effect of the OH-masks is required. However, some flux may be lost if the line is close to a portion of the OH-mask. For H$\alpha$ and [N~{\sc ii}]$\lambda$6583, we estimated this loss using Gaussian profiles at the appropriate emission line wavelength and line width. The average losses for H$\alpha$ and [N~{\sc ii}]$\lambda$6583 are 7.4\% and 13.4\%, respectively. The effect on the [N~{\sc ii}]$\lambda$6583/H$\alpha$ ratio is only 6.5\%, corresponding to $\sim0.016$ dex in 12+log(O/H), and thus no correction was applied. The composite spectra in the wavelength range around H$\alpha$ and [N~{\sc ii}]$\lambda\lambda$6548,6583 lines are presented in figure \ref{fig:SpectraHaN2}.

H$\alpha$ and [N~{\sc ii}]$\lambda\lambda$6548,6583 line fluxes are measured by spectral fitting of the stacked spectra with multi-Gaussian profile. The redshift and line width of other emission lines are based on that determined for H$\alpha$. We considered various ways of spectral stacking (standard mean, median, peak normalized) but little difference in the resulting line fluxes. The error on the metallicity derived from the stacked spectra is based on bootstrap resampling with 500 realizations. The metallicity in each bin is measured from the obtained stacked spectrum by using the N2 method described above, which is also summarized in table \ref{tab2}.

\subsection{[S~{\sc ii}]$\lambda\lambda$6716,6731 emission lines\label{subsec:s2lines}}
\begin{figure*}[]
\begin{center}
\includegraphics[angle=270, scale=1.00]{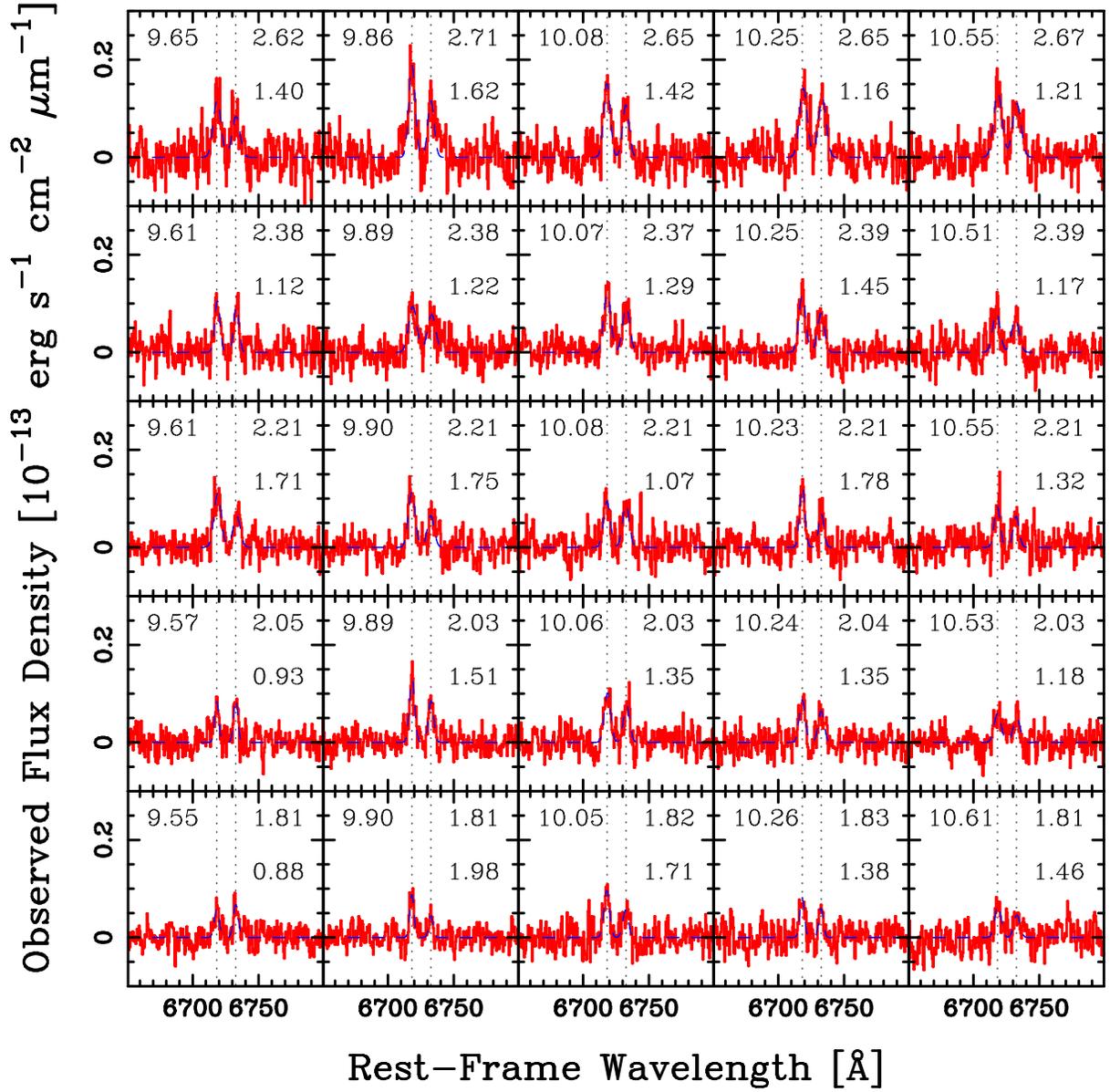}
\caption{Same as figure \ref{fig:SpectraHaN2}, but in the wavelength region around [S~{\sc ii}]$\lambda\lambda$6716,6731 lines, which are indicated by vertical dotted lines. The median values of log($M_{*}$[M$_{\solar}$]), log(SFR[M$_{\solar}$ yr$^{-1}$]), and [S {\sc ii}]$\lambda\lambda$6716,6731 line ratio are shown at the top-left, top-right, and middle-right part of each panel, respectively. \label{fig:SpectraS2}}
\end{center}
\end{figure*}

The wavelength range of our obtained spectra covers  [S~{\sc ii}]$\lambda\lambda$6716,6731 lines. The ratio of these two lines is sensitive to the electron density (e.g., \cite{Osterbrock:1989}). The composite spectra in the wavelength range around [S~{\sc ii}]$\lambda\lambda$6716,6731 lines are presented in figure \ref{fig:SpectraS2}, respectively. We have significant detections in the composite spectra. The line fluxes are measured by using the same spectral fitting as that for H$\alpha$ and [N~{\sc ii}] lines, which is also summarized in table \ref{tab2}. We discuss the electron density in subsection \ref{subsec:ED}.

\section{Results\label{sec:results}}
\subsection{The Mass-Metallicity Relation of Star-forming Galaxies at $z\sim1.4$\label{subsec:MZR}}

\begin{figure*}[]
\begin{center}
\includegraphics[angle=270, scale=0.50]{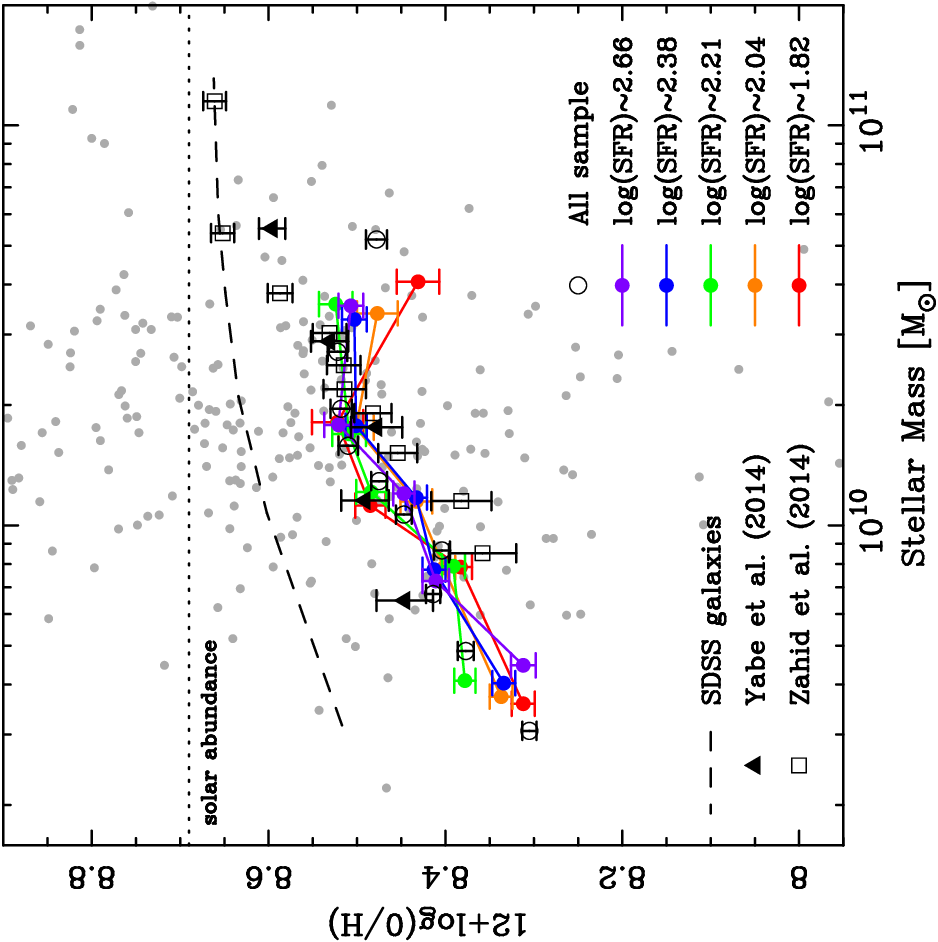}
\includegraphics[angle=270, scale=0.50]{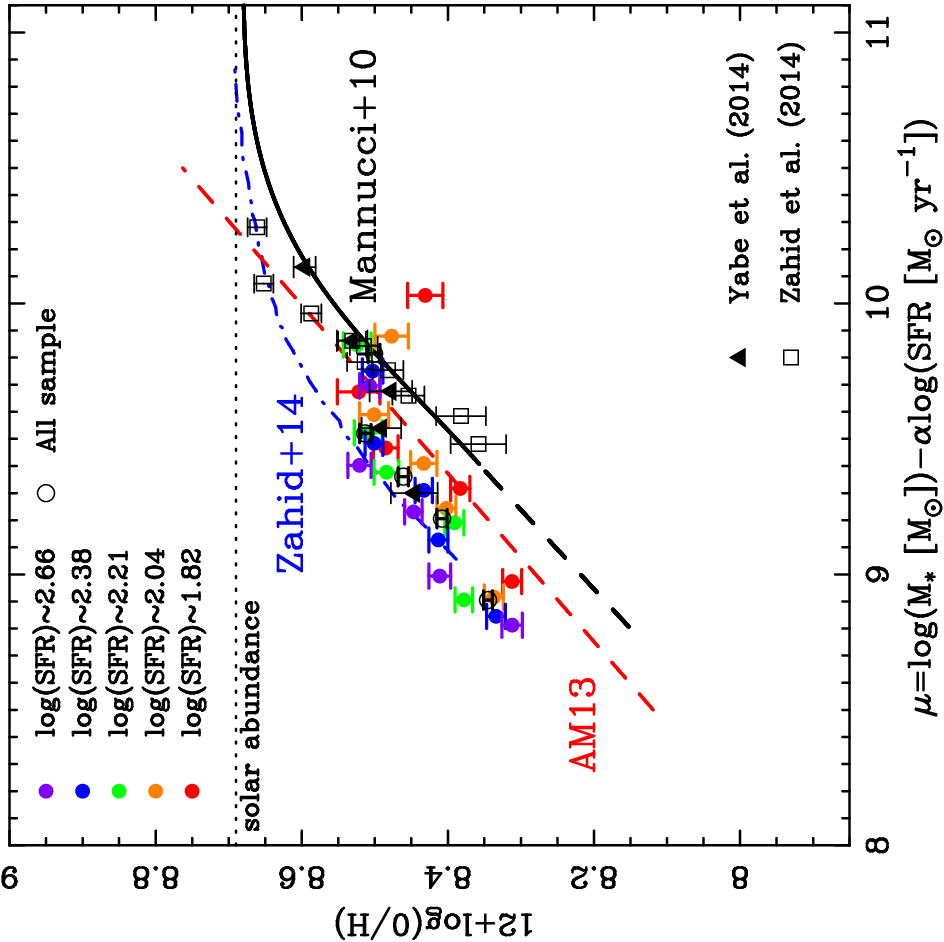}
\caption{(Left) The metallicity against the stellar mass. Individual object with both H$\alpha$ and [N~{\sc ii}] detection is indicated by a gray dot. The typical errors of the stellar mass and metallicity are 0.32 dex and 0.24 dex, respectively. The results from the stacked spectra in 10 stellar mass bins are indicated by open circle. The results from the stacked spectra in different SFR (in M$_{\solar}$ yr$^{-1}$) bins are shown by filled circles color-coded by the SFR. Filled triangles and open squares are results in the past FMOS observations by \citet{Yabe:2014} and \citet{Zahid:2014b}, respectively. For comparison, the result at $z\sim0.1$ obtained with the sample by \citet{Tremonti:2004p4119} by using the N2 method \citep{Erb:2006p4143} is shown as a dashed curve. (Right) The distribution of the metallicity against the projected axis $\mu$ of the FMR. Here, we use the projection parameter of $\alpha=0.32$ for our sample and that by \citet{Yabe:2014} and \citet{Zahid:2014b}, which is proposed for SDSS galaxies at $z\sim0.1$ by \citet{Mannucci:2010p8026}. For comparison, the results by \citet{Yabe:2014} and \citet{Zahid:2014b} are also presented as filled triangles and open squares, respectively. The FMR at $z\sim0.1$ by \citet{Mannucci:2010p8026} is presented by thick solid curve. For $\mu\lesssim9.4$, the solid curve is extrapolated linearly, which is shown by the thick dashed line. Two FMRs at $z\sim0.1$ by \citet{Zahid:2014b} and \citet{Andrews:2013} are presented by dot-dash curve and dashed line, respectively, where the SFR corrected for the SDSS fiber aperture effect is used. We use a projection parameter of $\alpha=0.32$ for these two samples. The metallicities are derived by using the N2 method. \label{fig:MZRFMR}}
\end{center}
\end{figure*}

The left panel of figure \ref{fig:MZRFMR} shows the distribution of the obtained metallicity and stellar mass at $z\sim1.4$. We also show the results from the stacking analysis with different color depending on the SFR. No clear SFR dependence of the mass-metallicity relation is seen in the left panel of figure \ref{fig:MZRFMR}. The mass-metallicity relation of our sample generally agrees with those of previous results with $\sim$340 galaxies at $z\sim1.4$ \citep{Yabe:2014} and $\sim$160 galaxies at $z\sim1.6$ \citep{Zahid:2014b}, where the spectra are taken with FMOS and the metallicity is derived with the same N2 method. For this comparison, we adopted a Salpeter initial mass function (IMF). In the massive end with $M_{*}\gtrsim10^{10.5}$ M$_{\solar}$, metallicities of our sample are lower than the earlier estimates by up to $\sim0.2$ dex. This may be partly due to the lack of red galaxies from our sample as we mentioned in section \ref{sec:sample}; galaxies at $z\sim1.4-1.6$ with the redder color and/or the larger color excess show the higher metallicity \citep{Yabe:2014, Zahid:2014b}.

In the left panel of figure \ref{fig:MZRFMR}, we compare our result at $z\sim1.4$ to that at $z\sim0.1$ obtained with the sample by \citet{Tremonti:2004p4119} but by using the N2 method \citep{Erb:2006p4143}. Here, stellar mass is converted to that obtained assuming Salpeter IMF. Our result and previous results by \citet{Yabe:2014} and \citet{Zahid:2014b} at $z\sim1.4-1.6$ show lower metallicities than the result at $z\sim0.1$ at a fixed stellar mass by up to $\sim0.2$ dex. The difference is generally larger in the lower stellar mass, which is similar to previous results at $z\sim1.4-1.6$.

There exist various uncertainties in the obtained mass-metallicity relation at $z\sim1.4$. As we mentioned in subsection \ref{sec:methods:mass}, the stellar mass of a large part of our sample is corrected because of the lack of the NIR/MIR data set. In addition to the typical correction factor of $\sim$0.2 dex, there is a scatter of $\sigma_{M_{*}}=0.22$ dex. There also exists the uncertainty in the  SFR estimation of $\sigma_{SFR}=0.26$ dex as described in subsection \ref{sec:methods:sfr}. We estimate the possible effect on the obtained mass-metallicity relation arose from these uncertainties. We resample randomly on the stellar mass vs. SFR plane in figure \ref{fig:MsSFR} assuming a Gaussian distribution with $\sigma_{M_{*}}$ and $\sigma_{SFR}$ mentioned above, and then apply the same stacking analysis describe in subsection \ref{sec:methods:metallicity}. We repeat this procedure 500 times. The standard deviation of the realized metallicities is $\sim0.03$ dex, which is almost similar to the error based on the bootstrap method as made in subsection \ref{sec:methods:metallicity}.

\begin{table*}
\caption{Emission line ratios, metallicity, N/O in each stellar mass and SFR bin of our sample.\label{tab2}}
\begin{center}
\begin{tabular}{ccccccc}
\hline
Bin & N2 index$^{a}$ & S2 index$^{b}$ & S2 ratio$^{c}$ & 12+log(O/H)$^{d}$ & log(N/O)$^{e}$ & 12+log(O/H)$_{corr}$$^{f}$ \\ & & & & & & \\
\hline
1 & -1.03$\pm$0.06 & -0.77$\pm$0.08 & 0.88$\pm$0.36 & 8.31$\pm$0.01 & -1.20$\pm$0.12 & 8.13$\pm$0.01 \\
2 & -0.91$\pm$0.04 & -0.81$\pm$0.05 & 1.98$\pm$0.55 & 8.38$\pm$0.01 & -0.98$\pm$0.08 & 8.23$\pm$0.01 \\
3 & -0.73$\pm$0.04 & -0.60$\pm$0.06 & 1.71$\pm$0.53 & 8.48$\pm$0.02 & -1.02$\pm$0.08 & 8.37$\pm$0.02 \\
4 & -0.66$\pm$0.07 & -0.76$\pm$0.07 & 1.38$\pm$0.46 & 8.52$\pm$0.03 & -0.74$\pm$0.12 & 8.43$\pm$0.03 \\
5 & -0.82$\pm$0.08 & -0.77$\pm$0.09 & 1.46$\pm$0.71 & 8.43$\pm$0.02 & -0.93$\pm$0.14 & 8.30$\pm$0.02 \\
6 & -0.99$\pm$0.06 & -0.84$\pm$0.06 & 0.93$\pm$0.29 & 8.34$\pm$0.01 & -1.04$\pm$0.10 & 8.15$\pm$0.01 \\
7 & -0.87$\pm$0.04 & -0.65$\pm$0.04 & 1.51$\pm$0.30 & 8.40$\pm$0.01 & -1.14$\pm$0.07 & 8.24$\pm$0.01 \\
8 & -0.82$\pm$0.05 & -0.62$\pm$0.05 & 1.35$\pm$0.29 & 8.43$\pm$0.02 & -1.11$\pm$0.08 & 8.28$\pm$0.02 \\
9 & -0.70$\pm$0.05 & -0.70$\pm$0.05 & 1.35$\pm$0.34 & 8.50$\pm$0.02 & -0.86$\pm$0.09 & 8.38$\pm$0.02 \\
10 & -0.74$\pm$0.06 & -0.73$\pm$0.09 & 1.18$\pm$0.52 & 8.48$\pm$0.02 & -0.87$\pm$0.13 & 8.34$\pm$0.02 \\
11 & -0.92$\pm$0.04 & -0.71$\pm$0.05 & 1.71$\pm$0.39 & 8.38$\pm$0.01 & -1.12$\pm$0.08 & 8.19$\pm$0.01 \\
12 & -0.89$\pm$0.04 & -0.74$\pm$0.05 & 1.75$\pm$0.47 & 8.39$\pm$0.01 & -1.06$\pm$0.08 & 8.21$\pm$0.01 \\
13 & -0.73$\pm$0.04 & -0.69$\pm$0.06 & 1.07$\pm$0.30 & 8.48$\pm$0.02 & -0.92$\pm$0.09 & 8.34$\pm$0.02 \\
14 & -0.69$\pm$0.04 & -0.69$\pm$0.06 & 1.78$\pm$0.49 & 8.51$\pm$0.02 & -0.85$\pm$0.08 & 8.37$\pm$0.02 \\
15 & -0.66$\pm$0.05 & -0.80$\pm$0.07 & 1.32$\pm$0.45 & 8.52$\pm$0.02 & -0.68$\pm$0.10 & 8.39$\pm$0.02 \\
16 & -0.99$\pm$0.06 & -0.89$\pm$0.06 & 1.12$\pm$0.33 & 8.33$\pm$0.01 & -0.99$\pm$0.10 & 8.12$\pm$0.01 \\
17 & -0.85$\pm$0.05 & -0.74$\pm$0.05 & 1.22$\pm$0.30 & 8.41$\pm$0.01 & -1.00$\pm$0.08 & 8.22$\pm$0.01 \\
18 & -0.82$\pm$0.04 & -0.69$\pm$0.04 & 1.29$\pm$0.26 & 8.43$\pm$0.01 & -1.02$\pm$0.06 & 8.25$\pm$0.01 \\
19 & -0.70$\pm$0.03 & -0.70$\pm$0.05 & 1.45$\pm$0.33 & 8.50$\pm$0.01 & -0.86$\pm$0.07 & 8.35$\pm$0.01 \\
20 & -0.70$\pm$0.04 & -0.81$\pm$0.08 & 1.17$\pm$0.44 & 8.50$\pm$0.01 & -0.71$\pm$0.10 & 8.35$\pm$0.01 \\
21 & -1.03$\pm$0.07 & -0.85$\pm$0.07 & 1.40$\pm$0.45 & 8.31$\pm$0.01 & -1.09$\pm$0.11 & 8.06$\pm$0.01 \\
22 & -0.86$\pm$0.05 & -0.77$\pm$0.05 & 1.62$\pm$0.37 & 8.41$\pm$0.01 & -0.97$\pm$0.08 & 8.19$\pm$0.01 \\
23 & -0.80$\pm$0.04 & -0.80$\pm$0.05 & 1.42$\pm$0.32 & 8.45$\pm$0.01 & -0.85$\pm$0.07 & 8.25$\pm$0.01 \\
24 & -0.66$\pm$0.04 & -0.70$\pm$0.04 & 1.16$\pm$0.21 & 8.52$\pm$0.02 & -0.82$\pm$0.06 & 8.35$\pm$0.02 \\
25 & -0.69$\pm$0.04 & -0.68$\pm$0.05 & 1.21$\pm$0.26 & 8.51$\pm$0.01 & -0.87$\pm$0.07 & 8.33$\pm$0.01 \\
\hline
\end{tabular}
\\Note -- The bin number corresponds to that in Table \ref{tab1}. Errors are estimated by using the bootstrap method. ({\it a}) log([N {\sc ii}]$\lambda$6583/H$\alpha$) ({\it b}) log([S {\sc ii}]$\lambda\lambda$6716,6731/H$\alpha$) ({\it c}) [S {\sc ii}]$\lambda6716$/[S {\sc ii}]$\lambda6731$ ({\it d}) Oxygen abundance determined from the N2 index ({\it e}) N/O abundance ratio determined from the N2S2 index ($\equiv$log([N {\sc ii}]$\lambda$6583/[S {\sc ii}]$\lambda\lambda$6716,6731)). ({\it f}) Oxygen abundance after correcting the effect of N/O abundance ratio.
\end{center}
\end{table*}

\subsection{The Fundamental Metallicity Relation\label{subsec:FMR}}
The right panel of figure \ref{fig:MZRFMR} shows the metallicity along the projected axis ($\mu=$ log($M_{*}$) $-$ $\alpha$log(SFR)) of the FMR defined by \citet{Mannucci:2010p8026}. Here, we use a projection parameter of $\alpha=0.32$, which minimizes the scatter of the local mass-metallicity relation. The metallicity of the sample at $z\sim0.1$ based on the calibration by \citet{Maiolino:2008} is converted to  [N~{\sc ii}]$\lambda$6583 to H$\alpha$ line ratio according to the \citet{Maiolino:2008} calibration, and then is converted to the metallicity again with the N2 calibration by \citet{Pettini:2004p7356}. The FMR at $z\sim0.1$ defined by \citet{Mannucci:2010p8026} is shown as thick solid curve, and the extension toward lower $\mu$,  based on a linear extrapolation, is shown as a dashed line.

While the SFR used in the FMR by \citet{Mannucci:2010p8026} is not corrected for the SDSS fiber aperture effect, the SFR in our sample is corrected for the FMOS fiber aperture effect as described in subsection \ref{sec:methods:sfr}. Recently, \citet{Andrews:2013} and \citet{Zahid:2014b} derived FMRs based on the SDSS sample by using SFRs corrected for the aperture effect. The resulting FMRs are shown in the right panel of figure \ref{fig:MZRFMR} by a dashed line and a dot-dash curve, respectively. Both results are based on the N2 method for the metallicity estimation. Although \citet{Andrews:2013} and \citet{Zahid:2014b} adopted a projection parameter of $\alpha=0.30$, which minimizes the scatter in metallicities of their mass-metallicity relations, here we use a projection parameter of $\alpha=0.32$ for comparison. The results change very little, even if we use $\alpha=0.30$. In the right panel of figure \ref{fig:MZRFMR}, we also show results from earlier works at $z\sim1.4-1.6$ by \citet{Yabe:2014} and \citet{Zahid:2014b}. All stellar masses and SFRs are based on the Salpeter IMF, and the metallicity calibration of the N2 method is assumed.

The right panel of figure \ref{fig:MZRFMR} shows that the metallicity of our sample is close to that of the local FMR by \citet{Mannucci:2010p8026}, where the SFR is not corrected for the aperture effect, at higher $\mu$. On the other hand, at lower $\mu$, the metallicity is larger than the local value by up to $\sim0.2$ dex. Our result at $z\sim1.4$ is located between the FMRs with aperture corrected SFR at $z\sim0.1$ by \citet{Andrews:2013} and by \citet{Zahid:2014b}.

\subsection{Electron Density\label{subsec:ED}}
\begin{figure}[!htb]
\begin{center}
\includegraphics[angle=270, scale=0.50]{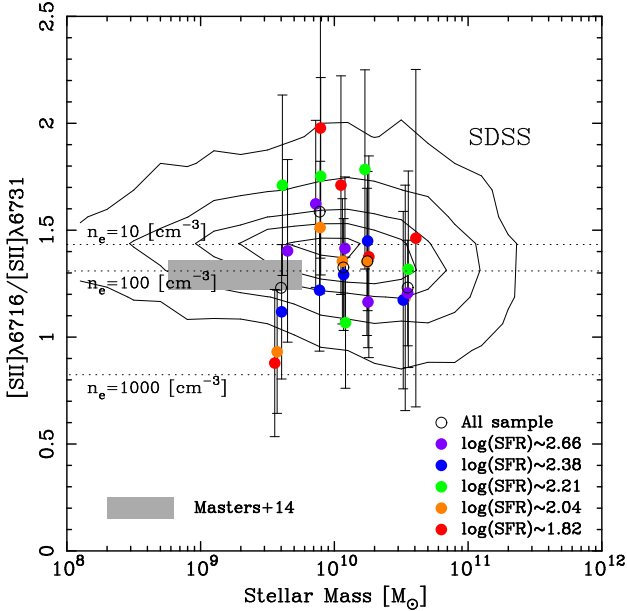}
\caption{The distribution of [S~{\sc ii}]$\lambda$6716/[S~{\sc ii}]$\lambda$6731 line ratio against the stellar mass. Symbols are color-coded in the same manner as figure \ref{fig:MZRFMR}. The contour shows the distribution of the local SDSS galaxies. The horizontal dotted lines are the corresponding electron densities. Here, we assume the electron temperature of 10,000 K. Gray shaded region indicates the results at $z\sim2$ by \citet{Masters:2014}. \label{fig:S2Ratio}}
\end{center}
\end{figure}

As presented in figure \ref{fig:SpectraS2}, we have significant detection of [S~{\sc ii}]$\lambda\lambda$6716,6731 doublet in the stacked spectra. The obtained [S~{\sc ii}]$\lambda$6716/[S~{\sc ii}]$\lambda$6731 line ratio ranges from $\sim1.0$ to $\sim1.8$ with a median value of 1.33. The corresponding electron density ranges $n_{e}\sim10$ to $\sim500$ cm$^{-3}$ with a median value of $\sim80$ cm$^{-3}$ assuming the electron temperature of 10,000 K \citep{Shaw:1995}. In this range of the line ratio, the electron density does not change significantly (up to a factor of 2) even if we assume the electron temperature of 100,000 K. The resulting line ratios, which are also summarized in table \ref{tab2}, are plotted against the stellar mass in figure \ref{fig:S2Ratio}. No clear trend between the electron density and other physical parameters such as stellar mass and SFR can be seen. The range of the line ratio of our sample is in good agreement with that of local galaxies selected from the SDSS DR8 catalogue \citep{Aihara:2011}, suggesting that galaxies at $z\sim1.4$ have similar electron density to local galaxies. The resulting electron density at $z\sim1.4$ roughly agrees with the recent result ($n_{e}=100-400$ cm$^{-3}$) of low mass galaxies at $z\sim2$ by \citet{Masters:2014}. 

\section{Discussion \label{sec:discussions}}


\subsection{Nitrogen-to-Oxygen Abundance Ratio \label{subsec:noo}}
There is a concern that the metallicity estimation has systematic errors (e.g., \cite{Nagao:2006, Kewley:2008}). The estimation based on the [N~{\sc ii}]$\lambda6583$ /H$\alpha$ line ratio depends on the ionization parameters (\cite{Denicolo:2002}; \cite{Kewley:2002}). Recently, \citet{Masters:2014} examine emission line galaxies at $z\sim2$, and find an offset from local galaxies on a [O~{\sc iii}]$\lambda\lambda4959,5007$/H$\beta$ vs. [N~{\sc ii}]$\lambda6583$ /H$\alpha$ diagram, also reported in earlier studies (e.g., \cite{Kewley:2013}), but no clear offset on a [O~{\sc iii}]$\lambda\lambda4959,5007$/H$\beta$ vs. [S~{\sc ii}]$\lambda\lambda$6716,6731/H$\alpha$ diagram. They conclude that the significant offset on the [O~{\sc iii}]$\lambda\lambda4959,5007$/H$\beta$ vs. [N~{\sc ii}]$\lambda6583$/H$\alpha$ diagram is caused by a nitrogen enhancement due to the high nitrogen-to-oxygen abundance ratio (hereafter N/O). \citet{Steidel:2014} suggest that the higher N/O ratio of high-redshift galaxies can partly explain a similar offset of high-redshift galaxies from local galaxies on the BPT diagram.  \citet{Hayashi:2015} also show that the N/O of galaxies at $z\sim1.5$ derived by using  [N~{\sc ii}]$\lambda6583$ and [O~{\sc ii}]$\lambda$3727 lines is  $>$ 0.1, which is higher than local galaxies at a fixed metallicity. In this subsection, we will examine the N/O of galaxies at $z\sim1.4$ using emission lines measured from the composite spectra of our large NIR spectroscopic sample.

\begin{figure}[!htb]
\begin{center}
\includegraphics[angle=270, scale=0.55]{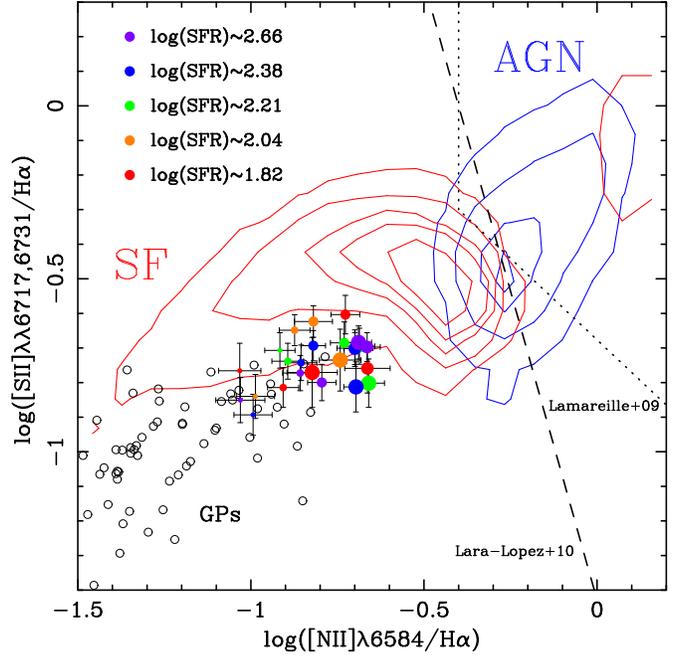}
\caption{[N~{\sc ii}]$\lambda$6583/H$\alpha$ versus [S~{\sc ii}]$\lambda\lambda$6716,6731/H$\alpha$. Our samples with different SFR groups are indicated as filled circles with the same color-coding as figure \ref{fig:MZRFMR}. The larger symbols show the higher stellar masses. The distribution of the SFGs and AGNs taken from the SDSS DR8 catalogue is shown by red and blue contour, respectively. The empirical lines separating SFGs and AGNs by \citet{Lamareille:2009} and \citet{Lara-Lopez:2010a} are shown by dotted and dashed line, respectively. The Green Pea galaxies (\cite{Cardamone:2009}, and \cite{Hawley:2012}) are indicated by open circles. \label{fig:N2S2}}
\end{center}
\end{figure}

Figure \ref{fig:N2S2} shows the distribution of [N~{\sc ii}]$\lambda$6583/H$\alpha$ vs. [S~{\sc ii}]$\lambda\lambda$6716,6731/H$\alpha$ in each stellar mass and SFR bin. Both line ratios are often used in diagnostics to distinguish between star-forming galaxies (SFGs) and AGNs (e.g., \cite{Osterbrock:1989}). Empirical separation lines between SFGs and AGNs on [N~{\sc ii}]$\lambda$6583/H$\alpha$ vs. [S~{\sc ii}]$\lambda\lambda$6716,6731/H$\alpha$ diagram, which are indicated in figure \ref{fig:N2S2}, have been proposed in previous studies (\cite{Lamareille:2009},  and \cite{Lara-Lopez:2010a}). Our data points are located in the region of the SFGs, which is sufficiently far from the AGN region, implying that the contamination of AGN to our sample would be very low. Note that broad line AGNs with the line width of $>1000$ km s$^{-1}$ are excluded from our sample beforehand. 

Figure \ref{fig:N2S2} shows that the distribution of our sample differs from that of SDSS normal SFGs: In our sample, both [N~{\sc ii}]$\lambda$6583/H$\alpha$ and [S~{\sc ii}]$\lambda\lambda$6716,6731/H$\alpha$ are lower than those of main part of the SDSS sample. In figure \ref{fig:N2S2}, our sample is compared to emission line galaxies at $z\sim0.1$ to $\sim0.3$ called Green Pea galaxies (GPs; \cite{Cardamone:2009}). Here we use the GP sample by \citet{Cardamone:2009} with the emission line fluxes summarized by \citet{Hawley:2012}. The GPs are generally located in the region with lower [N~{\sc ii}]$\lambda$6583/H$\alpha$ and [S~{\sc ii}]$\lambda\lambda$6716,6731/H$\alpha$ than the SDSS sample. In the high [N~{\sc ii}]$\lambda$6583/H$\alpha$ and [S~{\sc ii}]$\lambda\lambda$6716,6731/H$\alpha$ end, they overlap with our sample.

The correlation between [N~{\sc ii}]$\lambda$6583 to [S~{\sc ii}]$\lambda\lambda$6716,6731 line ratio and N/O has been investigated in local galaxies and empirical calibrations have been proposed (\cite{Perez-Montero:2009}, \cite{Amorin:2010}, and \cite{Hawley:2012}). In this work, we estimate the N/O ratio for our sample at $z\sim1.4$ by using the N2S2 index calibrated by \citet{Perez-Montero:2009}, where the N2S2 index is defined as follows: 
\begin{equation}
\textrm{N2S2}=\textrm{log}\left(\frac{I([\textrm{N\textsc{ii}}]\lambda6583)}{I([\textrm{S\textsc{ii}}]\lambda\lambda6716,6731)}\right).
\end{equation}
\citet{Perez-Montero:2009} investigate the relation between the N2S2 index and the N/O of local galaxies. Although the scatter of the relation is moderately large ($\sigma\sim0.3$ dex), they present the least-squares bisector linear fit:

\begin{equation}
\textrm{log(N/O)}=1.26 \times \textrm{N2S2}-0.86. \label{eq3}
\end{equation}

\begin{figure}[!htb]
\begin{center}
\includegraphics[angle=270, scale=0.50]{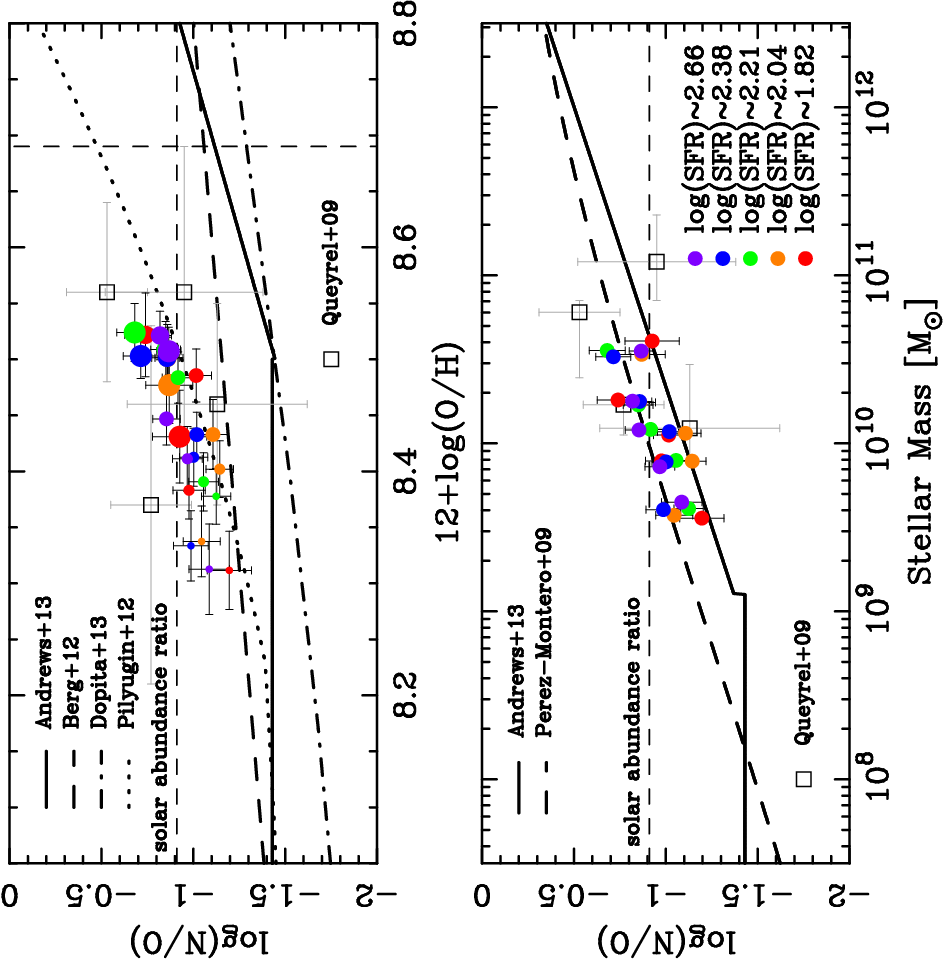}
\caption{(Top) N/O vs. 12+log(O/H). The symbols are the same as in figure \ref{fig:MZRFMR}. Galaxies with the larger stellar masses are indicated by the larger symbols. The thick solid, dashed, dot-dashed, and dotted lines are results for local galaxies by \citet{Andrews:2013}, \citet{Berg:2012}, \citet{Dopita:2013}, and \citet{Pilyugin:2012}, respectively. The open squares are results at $z=1.3-1.5$ by \citet{Queyrel:2009}. (Bottom) N/O versus stellar mass. Thick solid and dashed line are results by \citet{Perez-Montero:2009} and \citet{Andrews:2013}, respectively. The vertical and horizontal dashed line indicate the solar abundance and the solar abundance ratio, respectively \citep{Asplund:2009}. The open squares are results at $z=1.3-1.5$ by \citet{Queyrel:2009}. \label{fig:NoO}}
\end{center}
\end{figure}

In figure \ref{fig:NoO}, the N/O measured in each stellar mass and SFR bin of our sample galaxies at $z\sim1.4$ is presented as a function of metallicity and stellar mass, which is summarized in table \ref{tab2}. The mean (median) log(N/O) is $-0.95$ ($-0.98$). The N/O ratio increases with increasing both metallicity and stellar mass. The N/O ratio also appears to increase with increasing SFR at a fixed metallicity and stellar mass, although the trend is only marginal.

The obtained N/O of our sample at $z\sim1.4$ is compared with those of local galaxies in previous observations (\cite{Perez-Montero:2009}, \cite{Pilyugin:2012}, \cite{Berg:2012}, \cite{Andrews:2013}, and \cite{Dopita:2013}), where the stellar masses are scaled to those derived assuming the Salpeter IMF. The N/O at $z\sim1.4$ is generally higher than the local values at a fixed metallicity but almost comparable to those at a fixed stellar mass, which may be probably due to the difference of the mass-metallicity relation between local and high-redshift galaxies. The resulting N/O of galaxies at $z\sim1.4$ well agrees with those of massive and metal-rich part of the GPs sample by \citet{Hawley:2012}. It is worth noting that the deviation of N/O from local galaxies is significantly larger than the intrinsic uncertainty of the N/O calibration using the N2S2 index. 

\citet{Queyrel:2009} estimated the N/O of four star-forming galaxies at $z=$ 1.3 -- 1.5 by using the same N2S2 index. The obtained log(N/O) ranges from $-0.5$ to $-1.1$ in the metallicity range of 12+log(O/H) = 8.4 -- 8.6 and the stellar mass range of 10$^{10}$ -- 10$^{11}$ M$_{\solar}$, which are almost similar to those of our sample. More recently, the similarly high N/O of log(N/O) $\sim -1.0$ is also reported by using galaxy samples at $z\sim2$ (\cite{Masters:2014}, \cite{Steidel:2014}, and \cite{Hayashi:2015}).

One of the natural explanations of the variation in the N/O may be a time delay between the release of oxygen and that of nitrogen into the ISM; the two elements are formed in stars of different masses with different lifetimes (e.g., \cite{Edmunds:1978}, \cite{Garnett:1990}, \cite{Kobulnicky:1998}, and \cite{vanZee:1998}). In this scenario, after a single starburst, the N/O decreases and O/H increases due to the release of oxygen formed in massive, short-lived stars with the time-scale of a few tens of Myr. Then, the delayed release of nitrogen produced in intermediate mass, longer-lived stars increases N/O over longer time period of a few hundred Myr, while the O/H does not change significantly. If there are multiple starburst events, this cycle is repeated and both N/O and O/H increase gradually (e.g., \cite{Garnett:1990}, and \cite{Contini:2002}). The variance in the distribution of the N/O of local galaxies could be explained by the combination of this cycle. The high-redshift galaxies with high N/O may correspond to the upper locus of the distribution of local galaxies, where they are in the late quiescent phase of the nitrogen release. It may be unlikely, however, that all of the high-redshift galaxies are located in the quiescent phase.
 
IMF can change the chemical abundance ratio \citep{Kobulnicky:1998}. Since the nitrogen predominantly originates from low- or intermediate-mass stars, the N/O increases if the IMF has a steep slope or bottom-heavy shape. In recent studies, the evolution of the IMF with redshift has been suggested (e.g., \cite{vanDokkum:2008}). The slope of the massive end of the IMF at high redshift may depends on the physical properties of galaxies such as luminosity \citep{Hoversten:2008} and SFR \citep{Gunawardhana:2011}. The trend of the higher N/O at high redshift or high SFR regime, however, does not agree with the recent suggestions of flatter or top-heavy IMF.

The N/O enhancement by the nitrogen-rich (WN) Wolf-Rayet (WR) stars (e.g., \cite{Crowther:2007}) could be another cause of the higher N/O at high redshift (e.g., \cite{Kobulnicky:1998}, \cite{Henry:2000}, and \cite{Brinchmann:2008}). Increasing contribution of WR stars is expected in flatter IMF (e.g., \cite{Schaerer:1999}). In some studies of nitrogen enhanced galaxies, the spectroscopic features originated from WR stars have been reported (\cite{Pustilnik:2004}, \cite{Brinchmann:2008}, and \cite{Amorin:2012}). \citet{Perez-Montero:2011}, however, suggest no positive evidence of local pollution of nitrogen enriched gas from the WR stars in local blue compact dwarf galaxies with the integral field spectroscopy.

Another possible explanation is the dilution effect by the infall of metal poor gas (e.g., \cite{Edmunds:1990}). In this scenario, galaxies were originally more metal-rich and the distribution was close to the metal-rich part of local galaxies, where N/O increases with increasing metallicity because the secondary production of nitrogen is dominated (\cite{Koeppen:2005}, \cite{vanZee:2006}, and \cite{Masters:2014}). Then, the infall of the metal poor gas causes the decrease of the metallicity while N/O ratio does not change. In recent studies, especially in the theoretical context, gas inflow plays an important role in the galaxy growth. According to numerical models by \citet{Henry:2000}, however, the time required to approach the line of secondary production regime on 12+log(O/H) vs. N/O diagram is, at least, $>2$ Gyr, and thus, it may be unlikely that all of our sample galaxies at $\sim1.4$ were initially located in the secondary production regime of nitrogen (see discussion by \cite{Masters:2014}).

The differential outflow of oxygen-rich gas could be the origin of the higher N/O at high redshift. Recently, various studies suggest that the gas outflow is ubiquitous in star-forming galaxies at high redshift (e.g., \cite{Weiner:2009p13953}, \cite{Steidel:2010p10312}, and \cite{Yabe:2015}). Outflowing gas originated from supernova (SN) wind could be rich in oxygen relative to nitrogen because oxygen is generally formed in high-mass and short-lived stars while nitrogen is supplied from intermediate-mass and longer-lived stars. Although the gas outflow with well-mixed material does not change the abundance pattern of galaxies, the selective outflow of oxygen-rich gas driven by SN could cause the nitrogen enhancement \citep{vanZee:2006}. The direct observational evidence of the oxygen-rich gas outflow, however, would be desirable for further arguments.


\begin{figure*}[!htb]
\begin{center}
\includegraphics[angle=270, scale=0.50]{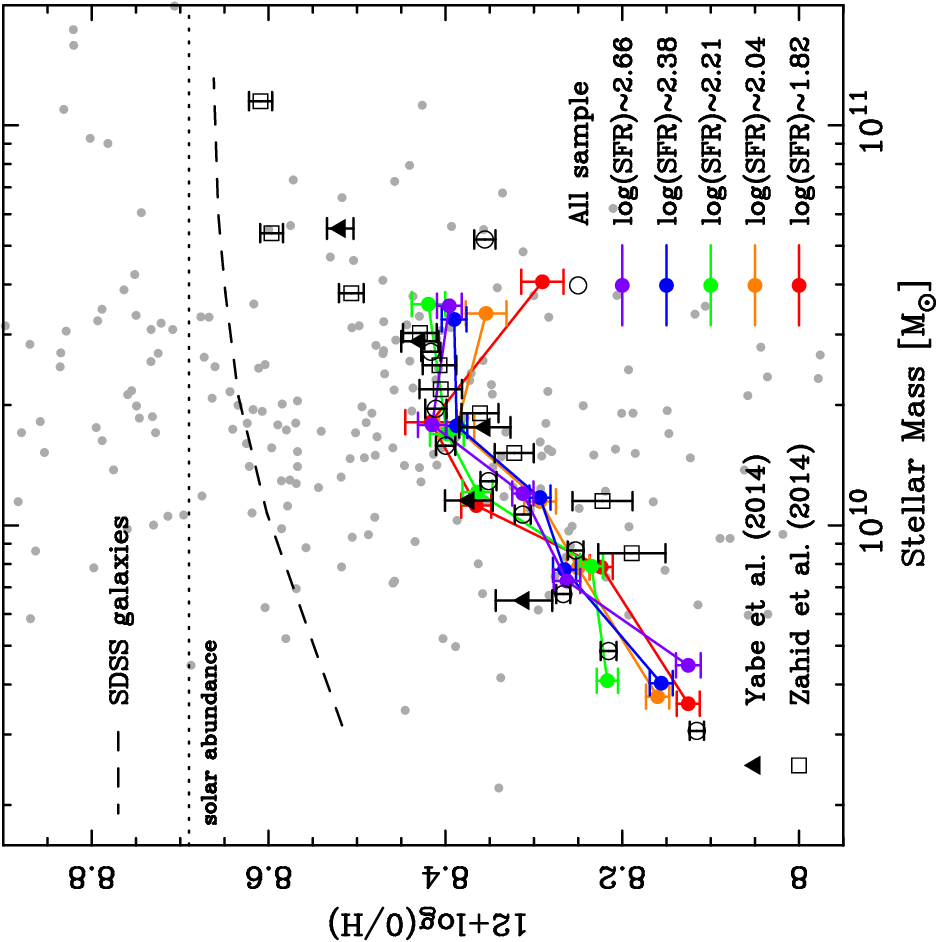}
\includegraphics[angle=270, scale=0.50]{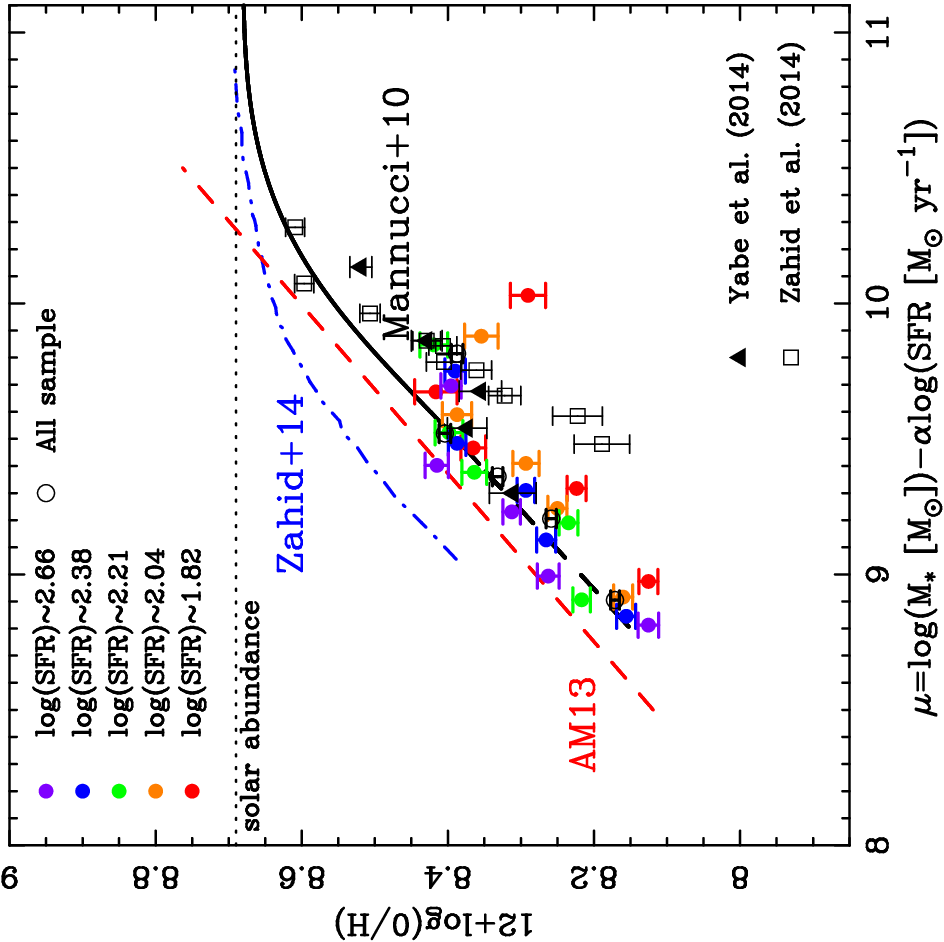}
\caption{The mass-metallicity relation (left) and the fundamental metallicity relation (right) based on the recalculated metallicity of our sample and those by \citet{Yabe:2014} and \citet{Zahid:2014b} corrected for the N/O enhancement. The symbols are all the same as figure \ref{fig:MZRFMR}.\label{fig:MZRFMRCorr}}
\end{center}
\end{figure*}

\subsection{The Mass-Metallicity Relation and the Fundamental Metallicity Relation Revisited \label{subsec:FMRR}}
In the previous  subsection, we show the higher N/O in high-redshift galaxies than local galaxies at a fixed oxygen abundance. This nitrogen enhancement causes a systematic effect on the metallicity estimation with the N2 method based on the [N~{\sc ii}]$\lambda$6583/H$\alpha$ flux ratio; if the N/O is higher than the local value, the metallicity derived from the N2 method is overestimated. \citet{Perez-Montero:2009} obtain the alternative N2 method taking into account the N/O effect:
\begin{equation}
\textrm{12+log(O/H)} = 0.79 \times \textrm{N2} - 0.56 \times \textrm{log(N/O)}+8.41, \label{Eq:N2NoOCorr}
\end{equation}
where N2 $=$ log([N~{\sc ii}]$\lambda$6583/H$\alpha$). This empirical calibration shows that the metallicity decreases with increasing N/O at a fixed N2 value. We recalculate the metallicity of our sample at $z\sim1.4$ by using the above calibration taking into consideration the N/O enhancement. We use the mean value of log(N/O) $=-0.95$ described in subsection \ref{subsec:noo} for all of our sample.

The metallicity corrected for the differing N/O is generally lower than the original metallicity by 0.1 -- 0.2 dex, which is also summarized in table \ref{tab2}. The left panel of figure \ref{fig:MZRFMRCorr} shows the distribution of the recalculated metallicity of our sample and previous results at $z\sim1.4-1.6$ by \citet{Yabe:2014} and \citet{Zahid:2014b} against the stellar mass. Here, the N/O ratio of previous studies is assumed to be the same value as our sample, and the metallicities are corrected in the same way as described above. The recalculated metallicities of our result are generally lower than that of the result at $z\sim0.1$ at a fixed stellar mass by up to $\sim0.4$ dex. The right panel of figure \ref{fig:MZRFMRCorr} shows the distribution of the corrected metallicity of our sample on the FMR diagram. Although there exists a discrepancy of up to $\sim0.3$ dex in the larger $\mu$, the result with the corrected metallicity agrees with the extrapolation of the local FMR by \citet{Mannucci:2010p8026}, where the SFR is not corrected for the aperture effect, in the other part. On the other hand, the resulting metallicity corrected for the N/O enhancement is lower than the results at $z\sim0.1$ with aperture corrected SFR by \citet{Andrews:2013} and \citet{Zahid:2014b} at fixed $\mu$, implying that the FMR may slightly evolve from $z\sim1.4$ to $z\sim0.1$.

\section{Summary \label{sec:summary}}
We present the results from a large near-infrared spectroscopic survey with Subaru/FMOS (\textit{FastSound}) consisting of $\sim$ 4,000 galaxies at $z\sim1.4$ with significant H$\alpha$ detection. We measure the gas-phase metallicity from the [N~{\sc ii}]$\lambda$6583/H$\alpha$ emission line ratio of the composite spectra in various stellar mass and star-formation rate bins. The resulting mass-metallicity relation generally agrees with previous studies obtained in a similar redshift range to that of our sample. The metallicity of our sample derived with the N2 method is lower than that at $z\sim0.1$ by up to 0.2 dex at a fixed stellar mass. No clear dependence of the mass-metallicity relation with star-formation rate is found. The metallicity of our sample is significantly larger than the FMR at $z\sim0.1$ derived by \citet{Mannucci:2010p8026} especially in lower $\mu$. The resulting metallicity roughly agrees with the FMRs at $z\sim0.1$ by \citet{Andrews:2013} and \citet{Zahid:2014b}, where the SFR is corrected for the fiber aperture effect and the metallicity is estimated with the N2 method. We detect significant [S~{\sc ii}]$\lambda\lambda$6716,6731 emission lines from the composite spectra. The electron density estimated from the [S~{\sc ii}]$\lambda\lambda$6716,6731 line ratio ranges from 10 -- 500 cm$^{-3}$, which generally agrees with that of local galaxies. On the other hand, the distribution of our sample on [N~{\sc ii}]$\lambda$6583/H$\alpha$ vs. [S~{\sc ii}]$\lambda\lambda$6716,6731/H$\alpha$ is different from that found locally, but is similar to that of lower redshift emission line galaxies such as Green Pea galaxies. We estimate the nitrogen-to-oxygen abundance ratio (N/O) from the N2S2 index, and find that the N/O in galaxies at $z\sim1.4$ is significantly higher than the local values at a fixed metallicity and stellar mass. The metallicity of our sample at $z\sim1.4$ recalculated with this N/O enhancement taken into account is lower than the original value by 0.1 -- 0.2 dex and is lower than the local value by up to 0.4 dex at a fixed stellar mass. The recalculated metallicity roughly agrees with the local FMR by \citet{Mannucci:2010p8026} but is lower than the FMRs with the aperture corrected SFRs by \citet{Andrews:2013} and \citet{Zahid:2014b}.

\bigskip
We thank Tohru Nagao and Matthew A. Malkan for helpful discussions. We are grateful to the FMOS support astronomer Kentaro Aoki for his support during the observations. The FastSound project was supported in part by MEXT/JSPS KAKENHI Grant Numbers 19740099, 19035005, 20040005, 22012005, and 23684007. KO is supported by the Grant-in-Aid for Scientific Research (C)(24540230) from the Japan Society for the Promotion of Science (JSPS). AB gratefully acknowledges the hospitality of the Research School of Astronomy \& Astrophysics at the  Australian National University, Mount Stromlo, Canberra where some of this work was done under the Distinguished Visitor scheme. This research has made use of the NASA/ IPAC Infrared Science Archive, which is operated by the Jet Propulsion Laboratory, California Institute of Technology, under contract with the National Aeronautics and Space Administration. We would like to express our acknowledgement to the indigenous Hawaiian people for their understanding of the significant role of the summit of Maunakea in astronomical research.



\end{document}